# Conversion of Legal Agreements into Smart Legal Contracts using NLP

Eason Chen
eason.tw.chen@gmail.com
Program of Learning Sciences
National Taiwan Normal University
Taipei, Taiwan

Niall Roche
n.roche@ucl.ac.uk
School of Management
University College London
London, UK

Yuen-Hsien Tseng
samtseng@ntnu.edu.tw
Graduate Institute of Library &
Information Studies
National Taiwan Normal University
Taipei, Taiwan

Walter Hernandez
walter.hernandez.18@ucl.ac.uk
Centre for Blockchain Technologies
University College London
and DLT Science Foundation
London, UK

Jiangbo Shangguan
jiangbo.shangguan@pku.org.uk
HSBC Business School
Peking University
Oxford, UK

Alastair Moore
a.p.moore@ucl.ac.uk
School of Management
University College London
London, UK

## ABSTRACT

A Smart Legal Contract (SLC) is a specialized digital agreement comprising natural language and computable components. The Accord Project provides an open-source SLC framework containing three main modules: Cicero, Concerto, and Ergo. Currently, we need lawyers, programmers, and clients to work together with great effort to create a usable SLC using the Accord Project. This paper proposes a pipeline to automate the SLC creation process with several Natural Language Processing (NLP) models to convert law contracts to the Accord Project's Concerto model. After evaluating the proposed pipeline, we discovered that our NER pipeline accurately detects CiceroMark from Accord Project template text with an accuracy of 0.8. Additionally, our Question Answering method can extract one-third of the Concerto variables from the template text. We also delve into some limitations and possible future research for the proposed pipeline. Finally, we describe a web interface enabling users to build SLCs. This interface leverages the proposed pipeline to convert text documents to Smart Legal Contracts by using NLP models.

## 1 INTRODUCTION

A Smart Legal Contract (SLC) is a specialized digital agreement consisting of natural language and computable components that both humans and machines can read. "The human-readable nature of the document ensures that signatories, lawyers, contracting parties, and others are able to understand the contract. The machine-readable nature of the document enables it to be interpreted and executed by computers, making the document 'smart' [15]". However, since SLCs are still in their early stages of development, lawyers, programmers, and clients need to work together while requiring significant effort to create them. Therefore, in this paper, we proposed a pipeline with a series of Natural Language Processing (NLP) models that can extract and transform the contents of a standard legal agreement into a SLC. Humans only need to correct the final output before using the converted SLC, thus lowering the barrier to SLC adoption.

In this paper, our main contributions are:

(1) We proposed a pipeline with several NLP models to convert raw Natural Language Legal Contracts into Smart Legal Contracts. The pipeline accurately detects CiceroMark from Accord Project templates with an accuracy of 0.8. Additionally, the Question Answering method can extract one-third of the Concerto variables.
(2) We examine the restrictions of the proposed pipeline when it comes to assessing existing data. Then, we proposed a direction for further research.
(3) We described an interface enabling users to build their Smart Legal Contracts.

## 2 RELATED WORK

The Accord Project provides an open-source SLC framework that contains three main modules: Cicero, Concerto, and Ergo. "These three modules are intertwined: Cicero utilizes Concerto to express variables in natural language that may be bound to Ergo for logic execution. [15]". The relationship between these modules is shown in Figure 1. An example contract is shown in Figure 2.

Cicero is the natural language template expression of the SLC, representing the contract text. Humans can read and edit Cicero to change the contract's variable elements, such as strings, dates, numbers and custom data types built on these primitives. Cicero's variable elements can be mapped to their corresponding type definitions in the Concerto data structure (Figure 2). This binding between the template elements and the data definition allows the contract to be more expressive and machine-readable than the natural language text alone. The Cicero and Concerto approach provides

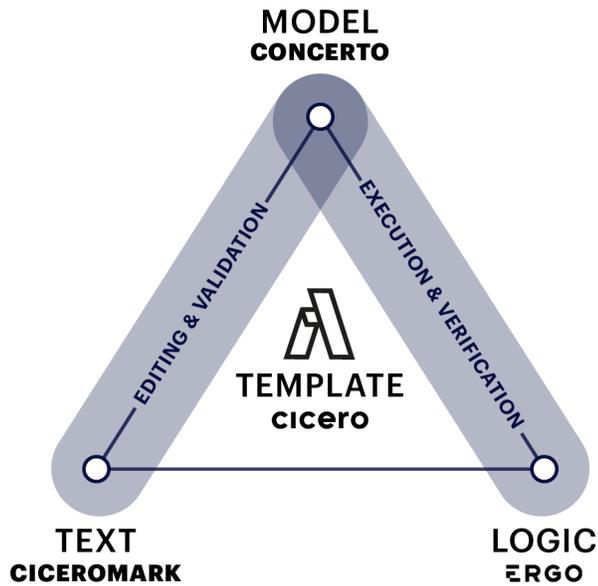

Figure 1: The relationship between Cicero, Concerto, and Ergo.

Cicero: Upon delivery and acceptance, {{buyer}} shall pay to {{seller}} the cost of goods ({{costOfGoods}}) and the delivery fee ({{deliveryFee}}).

Concerto:
```
{
  "buyer": "resource:org.accordproject.party.Party#BuyerName",
  "seller": "resource:org.accordproject.party.Party#SellerName",
  "costOfGoods": {
    "$class": "org.accordproject.money.MonetaryAmount",
    "doubleValue": 9.99,
    "currencyCode": "USD"
  },
  "deliveryFee": {
    "$class": "org.accordproject.money.MonetaryAmount",
    "doubleValue": 1.99,
    "currencyCode": "USD"
  }
}
```

Ergo:
```
emit PaymentObligation{
    contract: contract,
    promisor: some(contract.buyer),
    promisee: some(contract.seller),
    deadline: none,
    amount: contract.costOfGoods + contract.deliveryFee,
    description: toString(contract.buyer) ++
      " should pay cost of goods and delivery fee to "
      ++ toString(contract.seller)
};
```

Figure 2: Smart Legal Contract example: Payment Upon Delivery from the Accord Project's Template Library.

more structured contracts that are capable of being searched, analysed, and have improve the ability for supporting clauses that are capable of being executed.

Concerto is an object modeling language for defining and handling the data model of an SLC (Figure 2). Many pre-existing Concerto data types can be imported from the Accord Project's library[1].

[1]https://templates.accordproject.org

Specifically, the Concerto models can illustrate parties, transactions, and assets in an SLC agreement. In addition, Concerto can describe complex data types, such as new class declarations and their hierarchical relationship, by referencing and inheriting from other Concerto models. Concerto data models can be converted from/to JSON, and XML formats.

Ergo is a Domain Specific Language (DSL) that handles the execution of contract clause logic of the SLC [9, 15]. Ergo can be cross compiled into other programming languages for execution in a range of environments. When designing Ergo, the Accord Project intended it to be accessible to developers and lawyers. Developers can implement the logic of the SLC in Ergo, or other supported languages, such as TypeScript. Lawyers and developers can proofread the Ergo script to ensure consistency of function with the intent of the original legal clauses.

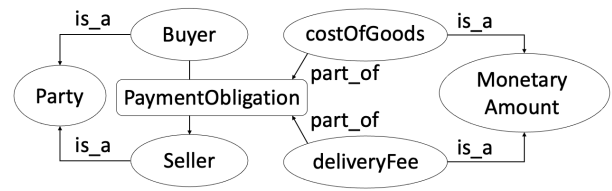

Figure 3: Knowledge Graph of the Payment Upon Delivery SLC constructed from its Cicero, Concerto, and Ergo.

Cicero, Concerto, and Ergo provide a structured representation of data that can be constructed into a Knowledge Graph, enabling the definition of entities and their relationships within the SLC. The data representation allows for complex relationships and types that support a hierarchy, facilitating the creation of more sophisticated smart legal contracts. For instance, in the Payment Upon Delivery example, Cicero indicates the location of the variables in the document, Concerto defines the type of variables such as "buyer" and "seller" is a "Party', while "costOfGoods" and "deliveryFee" is a "Monetary Amount" (see Figure 2 and Figure 3). In addition, Ergo captures the logical relationships between variables, such as specifying that the *buyer* should pay the *seller costOfGoods* and *deliveryFee* upon delivery and acceptance.

The focus of this paper is to convert plain text legal documents into CiceroMark [11] and Concerto [12] and then apply them to the existing Ergo template. Future work will explore generative approaches for producing Ergo for implementing contract logic that can process the CiceroMark and Concerto data.

## 3 METHODS TO CONVERT LEGAL CONTRACTS TO SLC BY NLP MODELS

To create an Accord Project SLC from a legal document, users must markup text on the raw text using CiceroMark variables and expressions, associate each variable with a corresponding data type using Concerto, and then write the executable aspects of the contract logic in Ergo. As this can be a complex and time consuming process, our work aims to improve SLC creation process by leveraging NLP technology with the following proposed steps.

## 3.1 Classify the Raw Contract Template

This step attempts to find a suitable template from similar SLC existing templates. Adapting other SLC similar contract or clause templates can speed up the creation process, since SLC with a similar purpose often shares the same or similar logic and data models. For example, a "Payment Upon Delivery" SLC data model (Figure 2) includes wording and variables to represent the buyer, seller, and price concepts that are largely similar to other contracts of the same type.

Our initial assumption was to classify the template using a supervised text classification approach, which inputs a contract text and outputs the most suitable category to which the template belongs. However, when uploading a new SLC Template, a pre-trained classification model must be fine-tuned. Aside from this constraint, since there are only 59 existing templates contributed to the Accord Project, it is highly likely that a pre-trained classification model would overfit.

As a result, we use the "more like this" query [6] similar search powered by Elasticsearch, which is a concept similar to KNN (k-nearest neighbors) where "category learning" is postponed until the classification phase [1]. When the user inputs the legal contract, the converter will list similar documents and their corresponding Cicero, Concerto, and Ergo implementation. After that, users can adapt these existing SLCs to build their ones. By doing so, a small SLC dataset can power the document classification process. Moreover, when users upload their new SLC to the SLC template library, we do not need to fine-tune the classification model because Elasticsearch only requires indexing to perform a similar search.

## 3.2 CiceroMark by Named Entity Recognition

Named Entity Recognition (NER) is an NLP task that locates and categorizes specific nouns from the raw text [16]. For example, the CoNLL-2003 English dataset marked named entities such as persons (PER), organizations (ORG), and locations (LOC), to name a few, from 909 documents [16]. We use NER techniques to mark potential text strings as Cicero variables and their corresponding data type from the raw text input (Figure 4). For example, we only need to mark an ORG or PER as a Party in the Concerto data model. We fine-tuned the RoBERTa pre-trained model [8] with the CONLL-2003 dataset [16] for NER tasks. We also included some aggregation of named entities in the training process. The table below shows examples of this aggregation:

| SLC Data type | From CONLL data types |
| --- | --- |
| B-String | All marks in the dataset |
| B-Party | B-per, B-org, B-geo, B-gpe |
| B-Object | B-art, B-MISC, B-nat |
| B-TemporalUnit | B-tim |

Table 1: The confusion matrix of the NER model on detecting Cicero Marks in Accord Project's templates.

When designing the NER model, we use multi-label classification [19]. We want the model to output several independent labels with individual probability instead of output results from a softmax layer. This is because some words in the contract might have multiple potential meanings [14], such as "London," which might represent both Party and Location. In such cases, we do not want the NER model to decide the final entity type, but rather, we want to let users edit it manually from the converter's interface.

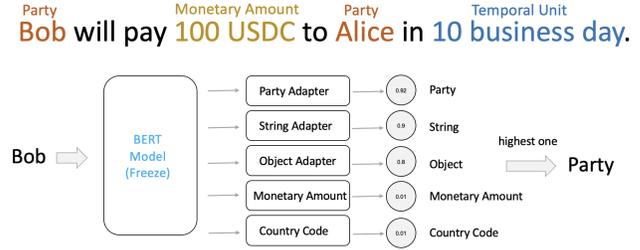

Figure 4: Example of NER with legal documents and how to leverage Adapter-Transformer to perform multi-class classification on each entity type in parallel

We implement the multi-label classification by using Adapter-Transformers [7, 10] on top of the RoBERTa based model [8] to provide several downstream models representative of each entity type's probability like Figure 4. The advantage of this approach is that the user can customize new data types and specify the training data to fit their domain. Furthermore, since the adapter's training is relatively fast [7] the converter can retrain labels dynamically when new data is added. Lastly, because the adapter model size is tiny [7], the converter can record each adapter version and note which tags and their versions were used in each conversion so that the user can reproduce the result in the future.

## 3.3 Concerto Data by Question Answering

Question Answering (QA) is a task that requires a machine learning model to perform reading comprehension ability [13]. The user provides a question and a context string to a QA model, which searches for an answer within the given context. Then, the QA model can output the location index span of the correct answer from the context.

We use the pre-trained QA model (deepset/roberta-base-squad2) [2] to extract the data from a legal contract to the Concerto Data model. If we ask the question using correct question words, the performance is impressive. For example, given an SLC contract and its Concerto data model below:

Bob will be deemed to have completed its delivery obligations if in Alice's opinion, the Widgets satisfies the Acceptance Criteria, and Alice notifies Bob in writing that she is accepting the Widgets.

> {
> "shipper": "Bob",
> "receiver": "Alice",
> "deliverable": "Widgets",
> }

If we ask the QA model with only a word like "deliverable" directly when extracting the deliverable from a contract, the model will output a sentence that contains the answer.

> **Q**: 'deliverable?'
> **A**: 'Bob will be deemed to have completed its delivery obligations if in Alice's opinion, the
Widgets' (**Confidence: 0.4**)

However, if we add some question words before the keyword like below, the model will provide a more precise answer:

**Q**: 'What is the deliverable?'

**A**: 'the Widgets' (**Confidence: 0.7**)

If we append the requirement to the model and state that we do not want 'the' in the answer, the model can understand and provide the correct answer we want.

**Q**: 'What is the deliverable without the?'

**A**: 'Widgets' (**Confidence: 0.6**)

Furthermore, if we use different question word, such as "where", we might get a completely different answer.

**Q**: 'Where is the deliverable?'

**A**: 'Bob' (**Confidence: 0.6**)

The confidence score was calculated by the mean of the models' confidence in the answer's start and end indexes. Due to the limit of a BERT-like model [5], when processing a document longer than 512 tokens, we will split the document into several parts with some overlap and send them to the QA model, and then pick the answer with the highest confidence.

### 3.4 Output and Further Conversion

The output from the converter is the Cicero file with CiceroMarks and Concerto data models in JSON format, which can then use Accord Project's converter to generate the Concerto of the SLC. As a result, users will gain a nearly ready-to-use SLC, with a reduction of manual effort. Users may still need to adapt or extend the contract with the appropriate Ergo logic. However, the nearly ready-to-use SLC is a starting point that should be relevant for them to base their implementation on.

## 4 EVALUATION AND DISCUSSION

The Accord Project's template library comprises 56 SLC templates, consisting of their corresponding Cicero, Concerto, and Ergo components. These templates are designed to standardize data types, individual clauses, and complete contracts for reuse and extension. After filtering by version and availability, we selected 53 of them. After filtering based on version and availability, we have selected 53 templates for use in evaluating the performance of our proposed conversion process.

### 4.1 CiceroMark by Named Entity Recognition

We utilized Named Entity Recognition (NER) with proposed models to detect CiceroMarks. We used the model RoBERTa-base [8], fine-tuned with Adapter-Transformer [7, 10] for each label with batch size 256 for three epochs. We consider a token to have a positive label if the NER model has identified it as anything other than the "O" label. The general accuracy of the result was 0.81. However, the general F1 score was only 0.32, indicating a significant imbalance in the dataset. From the confusion matrix (Table 2), we observed that out of 41973 labels, 32013 data were correctly labeled as negative (O-class), while 1839 data were correctly labeled as positive. Notably, there were 7482 false positives and 639 false negatives. Notably, there were 7482 false positive and 639 false negative.

| Actual\Predicted | True | False |
|---|---|---|
| True | 1839 | 639 |
| False | 7410 | 32085 |

Table 2: The confusion matrix of the NER model on detecting Cicero Marks in Accord Project's templates.

We further dived into the false positive data, like the example on Figure 5, and discovered that most of these data commenced with capital letters and were categorized as Party or Object. This was because the templates capitalized many keywords but did not mark them as CiceroMarks, as demonstrated in the provided example. Additionally, we found that the model needed to improve handling addresses, email, and currency codes, likely due to the limited training data in this domain, resulting in many false negatives.

After execution of this Agreement at 50 Bidborough Street, London, "Dan" shall pay the full purchase price to "Jerome" in the amount of 3.14 EUR upon demand by "Jerome".

Figure 5: Example output of NER model with correct and incorrect labels.

Furthermore, upon dividing the performance by templates, we observed 22 templates from 56 with an F1 score exceeding 0.7 (Figure 6). This observation highlights that our model can accurately identify CiceroMarks in specific pre-existing templates.

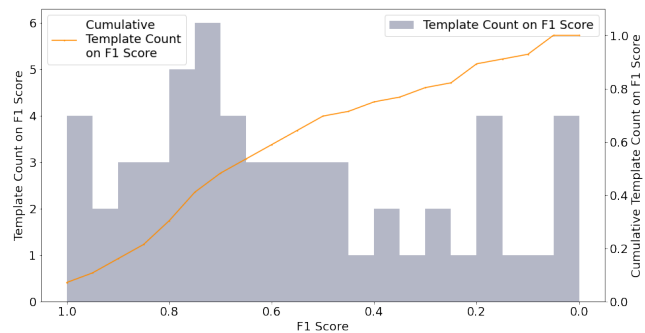

Figure 6: The distribution and accumulation of the models' performance by Templates from the Accord Project.

### 4.2 Extract Concerto Data with Question Answering

To extract data from a legal contract and map it to the Concerto Data model using Question Answering, we employed some popular pre-trained QA models, including "deepset/roberta-base-squad2" [2], "deepset/bert-base-cased-squad2" [4], "distilbert-base-cased-distilled-squad" [17], and "bert-large-uncased-whole-word-masking-finetuned-squad" [5]. Based on the results in Table 3, "deepset/roberta-base-squad2" [2] displayed the best performance, with an F1 score of 0.395. Although, the F1 score is low, the 0.325 exact match (EM) rate implies that the model can accurately extract one-third of the template data perfectly.

| Model Name | F1 | EM |
| --- | --- | --- |
| deepset/roberta-base-squad2 | 0.395 | 0.325 |
| deepset/bert-base-cased-squad2 | 0.185 | 0.128 |
| distilbert-base-cased-distilled-squad | 0.173 | 0.128 |
| bert-large-uncased-whole-word-masking-finetuned-squad | 0.318 | 0.262 |

Table 3: The performance of different QA pre-trained models.

We also experimented with different question prefixes and discovered that asking questions beginning with "what" achieved the highest general F1 and EM scores (Table 4). Although, some variables may perform better with other interrogative sentences, such as "how much" will do better for numbers and "where" will do better for address, we use "what" as the prefix for all questions in this research.

| Interrogative sentences | F1 | EM |
| --- | --- | --- |
| What is the (variable_name)? | 0.395 | 0.325 |
| Where is the (variable_name)? | 0.348 | 0.271 |
| How much is the (variable_name)? | 0.272 | 0.204 |
| How is the (variable_name)? | 0.270 | 0.197 |
| When is the (variable_name)? | 0.252 | 0.193 |
| (variable_name)? | 0.307 | 0.239 |
| (variable_name) | 0.215 | 0.176 |

Table 4: The data combination for SLC data types from CONLL data types.

Furthermore, if we ignored low-confidence answers, we noticed a significant increase in the EM rate (Figure 7). For example, if the confidence criteria are set at 0.3, the EM rate for the remaining data will increase to 0.6 from 0.32.

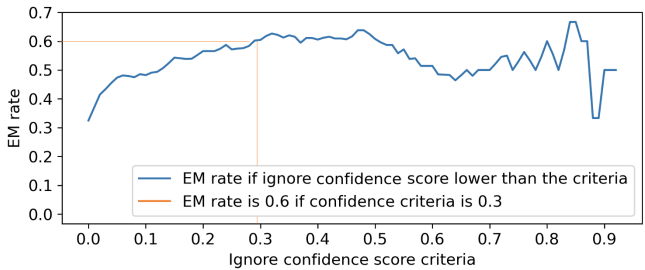

Figure 7: EM rate if we ignored low-confidence answers.

During this stage, we also identified certain limitations in the evaluation process. One limitation is that if the correct answer to a monetary amount question is "500.00" and the QA model's answer is "500", the F1 and EM scores will be calculated as 0. Moreover, if the answer is "500 USD" and the QA model's answer is "500", the F1 score will be 0.67, and the EM will be 0.

## 5 POTENTIAL USER FLOW FOR THE CONVERTER

In this section, we will outline a proposed user flow and interface to help users create SLC using the NLP converter discussed in this paper. However, based on the findings of the previous section, it is clear that the pipeline alone is not sufficient to effectively convert legal documents to SLC. Despite this, the pipeline may still be useful in assisting human users in creating their own SLC, provided they can review and correct its outputs [3]. Therefore, the proposed user flow consists of the following steps:

(1) Upload the legal documents.
(2) Select a similar SLC Template.
(3) Run the model to mark CiceroMarks by NER model.
(4) Filling the Concerto data model by QA model.
(5) Correct the output and customize the data model.
(6) Convert the output as Accord Projects' SLC.

First, users need to upload the legal document they want to convert. Then, the converter will show similar templates for users to apply. As Figure 8 shows below:

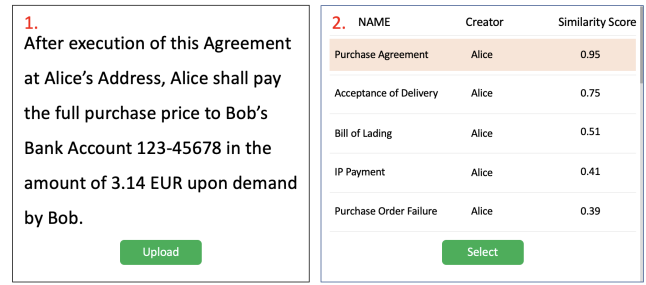

Figure 8: User interface to upload a contract and select a suitable SLC template.

Next, the SLC converter will load the selected template to perform NER, highlighting all potential keywords in various colors based on their entity type, which represent potential position for Cicero variables. Users can fine-tune the selection or deselection of the CiceroMarks or customize the version of the NER Adapter as necessary (Figure 9). Finally, users can manually input the legal contract's data to the Concerto data model or use the QA model to extract and revise each section's information (Figure 9).

Moreover, suppose users consider the data model does not fit their needs and want to modify it. They can use the Accord Project's Concerto Data Model Editor to build their own custom Concerto data model (Figure 10).

Eventually, users can optionally contribute this newly created SLC to the Accord Project's Template Library. By doing so, the converter's NLP model can perform the Active Learning method [18]. After training the new version's models based on the data feedback from the users, the SLC converter will remain up-to-date for the subsequent conversion processes.

## 6 CURRENT RESULTS AND FUTURE WORKS

The NLP models and the active learning retrain pipeline proposed in this paper are available via a REST API. The performance of these models on some existing SLC templates is acceptable. However, due to data limitations, we are unable to employ a proper validation method, such as cross-validation, to test our pipeline. The Accord Project's template library only consists of 59 templates, each designed for a unique purpose, making it difficult to form a comprehensive validation set. Therefore, we are building the interface present in Section 5. By doing so, we enable the public to contribute new templates with the aid of NLP models from our proposed pipeline. We can then leverage user feedback to improve the performance of these NLP models.

Figure 9: User interface to edit Cicero and Concerto data models.

Figure 10: Screenshot of the Accord Project's Concerto Data Model Editor component.

Second, we might generate new data to train the NLP models. We use the CoNLL-2022 dataset [16] to train the NER Adapters to generate CiceroMarks. However, the CoNLL-2022 dataset [16] lacks some named entities which are often used in the legal contract domain, such as Time Zone, Currency Code, and Address. Therefore, we will perform data augmentation to create more data from the same template to enhance the NER model's performance. Moreover, we will use generative models such as ChatGPT to produce more templates for training and evaluation.

Third, when using Adapter-Transformer [10] to perform multi-label NER, we found that the time cost to run the model increases linearly with the number of Adapters we use. Hence, we should consider using other approaches to implement the multi-label classification to speed up the process while keeping the model lightweight and portable in the future.

Finally, based on the findings of this research, we plan to use a "question generator" to determine the appropriate question prefix required before the QA model extracts information from the input. Hence, the converter can extract accurate information from legal contracts and map it to the Accord Project's Concerto data model.

## 7 CONCLUSION

We proposed a pipeline with several NLP models to convert raw natural language contracts to Smart Legal Contracts. After evaluating the proposed pipeline, we discovered that our NER pipeline accurately detects CiceroMarks [11] from the Accord Project templates' text with an accuracy of 0.8. Additionally, our Question Answering method can successfully extract one-third of the Concerto variables from the template text. We also discuss some limitations and possible future work for the proposed pipeline. Finally, we describe an interface enabling users to build their SLC with the proposed pipeline to convert text documents to Smart Legal Contracts with the help of NLP models.

## ACKNOWLEDGMENTS

This work received support from Google Summer of Code and was partially supported by the Ministry of Science and Technology of Taiwan (R.O.C.) under Grant 109-2410-H-003-123-MY3. We express our gratitude to the Accord Project's members, particularly, Matt Roberts, Jérôme Siméon, and Dan Selman for their assistance with this project.